\documentclass{PoS}
\usepackage[numbers]{natbib}
\pdfoutput=1

\title{Cygnus X-1: shedding light on the spectral variability of a black
  hole}

\ShortTitle{Cygnus X-1: shedding light on the spectral variability of
  black holes}

\author{
V.\,Grinberg$^{a}$,
  N.\,Hell$^{a,b}$,
  J.\,Wilms$^{a}$,
  J.\,Rodriguez$^{c}$,
K.\,Pottschmidt$^{d,e}$,
M.A.\,Nowak$^{f}$,
M.\,B\"ock$^{g}$,
A.\,Bodaghee$^{h}$,
M.\,Cadolle Bel$^{i}$,
F.\,F\"urst$^{a,j}$,
M.\,Hanke$^{a}$,
M.\,K\"uhnel$^{a}$,
P.\,Laurent$^{k}$,
S.B.\,Markoff$^{l}$,
A.\,Markowitz$^{m}$,
D.M.\,Marcu$^{d,e}$,
G.G.\,Pooley$^{n}$,
A.\,Popp$^{a}$,
R.E.\,Rothschild$^{m}$,
J.A.\,Tomsick$^{h}$\\
\llap{$^{a}$} Remeis-Observatory/ECAP/FAU, Bamberg, Germany\\
\llap{$^{b}$} LLNL, Livermore, CA, USA\\
\llap{$^{c}$} Lab. AIM, CEA-Saclay, France\\
\llap{$^{d}$} CRESST/NASA-GSFC, Greenbelt, MD, USA\\
\llap{$^{e}$} UMBC, Baltimore, MD, USA\\
\llap{$^{f}$} MIT, Cambridge, MA, USA\\
\llap{$^{g}$} MPIfR, Bonn, Germany\\
\llap{$^{h}$} SSL, UC Berkeley, Berkeley, CA, USA\\
\llap{$^{i}$} ESAC, Madrid, Spain\\
\llap{$^{j}$} SRL/Caltech, Pasadena, CA, USA\\
\llap{$^{k}$} APC, Univ. Paris Diderot, CNRS/IN2P3, CEA/Irfu, Obs. de Paris, Sorbonne Paris Cité, France\\
\llap{$^{l}$} UvA, Amsterdam, The Netherlands\\
\llap{$^{m}$} CASS/UCSD, La Jolla, CA, USA\\
\llap{$^{n}$} University of Cambridge, Cambridge, UK\\
}

\abstract{ The knowledge of the spectral state of a black hole is
  essential for the interpretation of data from black holes in terms
  of their emission models. Based on pointed observations of
  \mbox{Cyg\,X-1} with the Rossi X-ray timing Explorer (\textsl{RXTE})
  that are used to classify simultaneous \textsl{RXTE}-ASM
  observations, we develop a scheme based on \textsl{RXTE}-ASM colors
  and count rates that can be used to classify all observations of
  this canonical black hole that were performed between 1996 and
  2011. We show that a simple count rate criterion, as used
  previously, leads to a significantly higher fraction of
  misclassified observations.  This scheme enables us to classify
  single \textsl{INTEGRAL}-IBIS science windows and to obtain summed
  spectra for the soft, intermediate and hard state with low
  contamination by other states.}

\FullConference{"An INTEGRAL view of the high-energy sky (the first 10
  years)"
  9th INTEGRAL Workshop and celebration of the 10th anniversary of the launch,\\
  October 15-19, 2012\\
  Bibliotheque Nationale de France, Paris, France}

\begin{document}

\section{The States of Black Hole Binaries}

\begin{figure}
\begin{minipage}{0.58\textwidth}
\includegraphics[width=\textwidth]{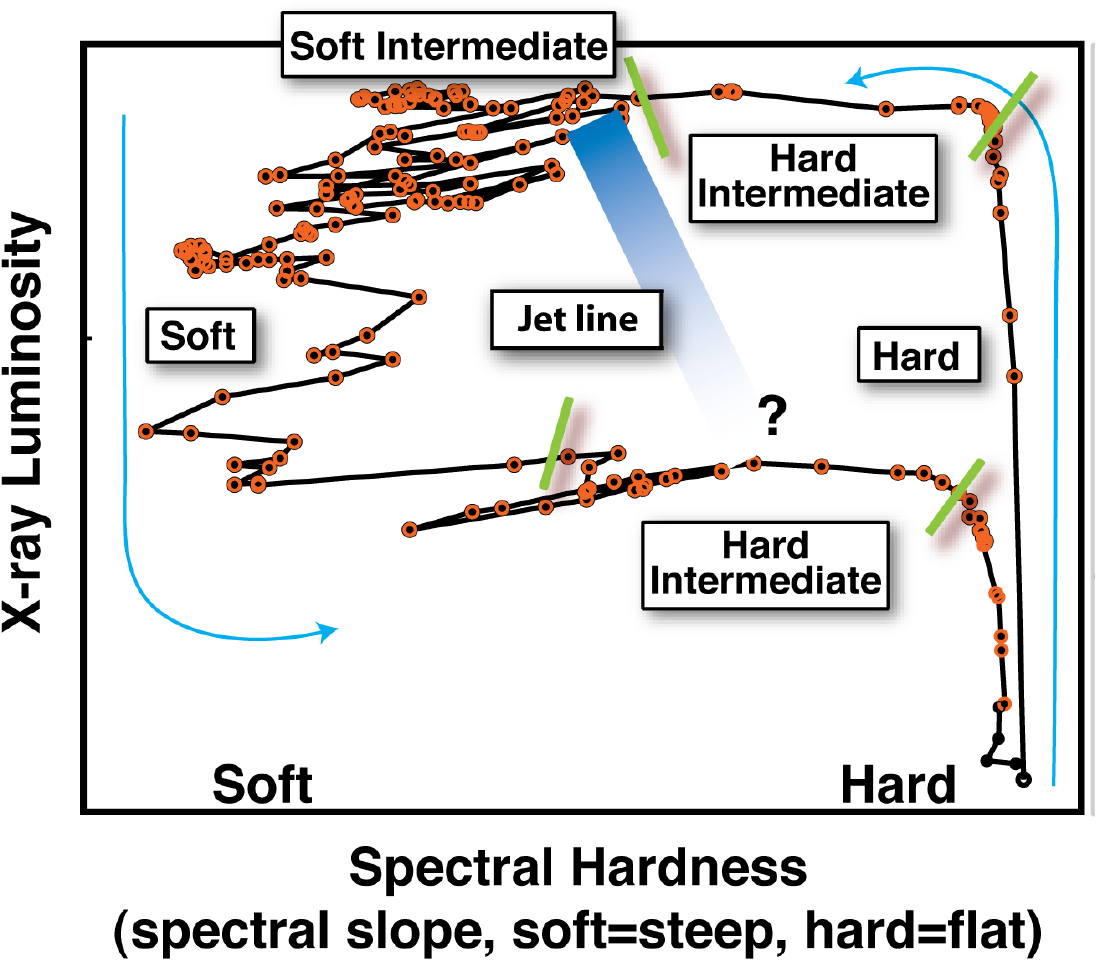}
\end{minipage}\hfill
\begin{minipage}{0.38\textwidth}
  \caption{The typical q-track of transient black hole binaries. From:
    \texttt{http://www.sternwarte.uni-}\newline\texttt{erlangen.de/proaccretion/}}\label{fig:q}
\end{minipage}
\end{figure}

Transient black hole binaries (BHBs) in outburst move on
characteristic q-shaped tracks through the hardness-intensity-diagram
(HID, Fig.~\ref{fig:q}): coming from quiescence a source enters the
hard state, which is followed by an intermediate state (subdivided
into hard intermediate (HIMS) and soft intermediate
(SIMS))~and~finally the soft state.  Then the source returns into a
hard state, albeit usually at lower luminosities than previously and
finally into quiescence (for a detailed discussion of states see
\cite{Fender_2004a,Fender_2009a,Belloni_2010a}).  The individual
states are also clearly different in their timing and radio
properties. The fact that the soft states show, as opposed to the hard
states, no or only quenched radio emission, implies the absence or
weakness of the jet. The HIMS to SIMS transition happens close to or
at the jet line, where the properties of the jet and therefore the
geometry of the source change \cite{Fender_2009a}.

\section{Cygnus X-1}

\begin{figure}
\includegraphics[width=\textwidth]{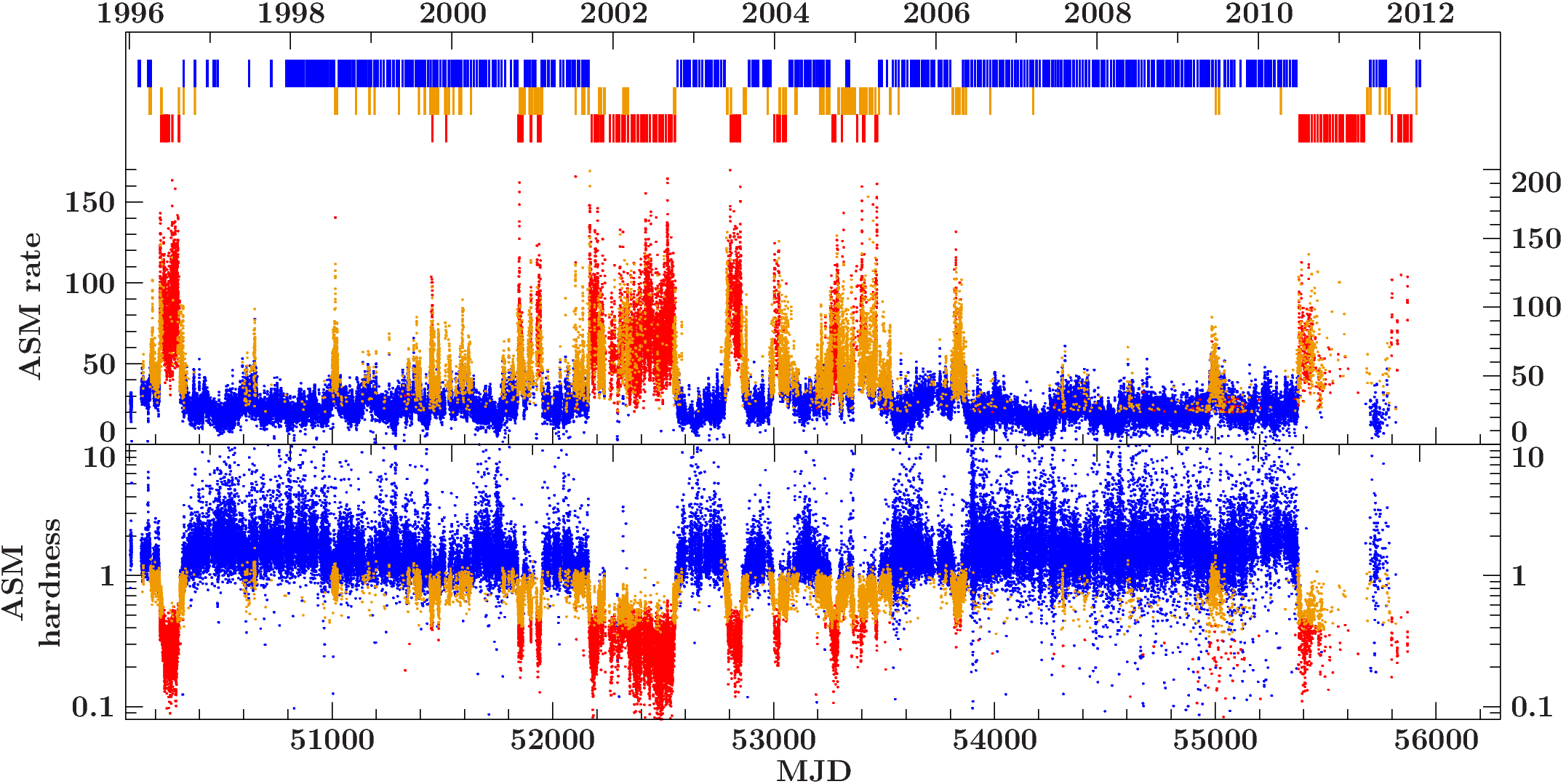}
\caption{ \textsl{RXTE}-ASM light curve of \mbox{Cyg\,X-1}. Dashes
  represent times of pointed \textsl{RXTE} observations, classified
  into states (blue: hard, orange: intermediate, red: soft) using the
  $\Gamma_1$-based state definition introduced in
  Sect.~\protect\ref{sec:states}. The individual ASM measurements are
  classified into different states (same color code as for
  $\Gamma_1$-based state definitions) according to the ASM-based
  classification introduced in Sect.~\protect\ref{sec:states} and on
  Fig.~\protect\ref{fig:map}.}
\label{fig:monster}
\end{figure}

\mbox{Cyg\,X-1} is a bright persistent black hole binary that often
undergoes (failed) state transitions \cite{Pottschmidt_2003b}. That
makes it a prime target for both long-term monitoring (e.g., with
\textsl{RXTE}, \textsl{INTEGRAL} or Ryle/AMI) and snapshot
observations (e.g., with \textsl{XMM}, \textsl{Chandra} or
\textsl{Spitzer}). \textsl{RXTE} has observed \mbox{Cyg\,X-1} from
1996 to the end of mission in late 2011, largely as a part of our
bi-weekly campaign \cite{Pottschmidt_2003b,Wilms_2006a}.  During this
time the source has undergone periods with different source activity
patterns, e.g., the long very hard state from mid-2006 to mid-2010
(Fig.~\ref{fig:monster}). State transitions are self-similar on
different timescales and can happen as quickly as within a few hours
\cite{Boeck_2011a}.

We extract all available PCA and HEXTE data of the source for every
\textsl{RXTE} orbit and obtain 2741 individual spectra which we model
with a combination of a broken power law (with a soft photon index
$\Gamma_1$, hard photon index $\Gamma_2 < \Gamma_1$, and a break
energy $E_{\mathrm{break}} \sim 10$\,keV), an high energy cut off, an
iron line and, where required, a thermal disk
component. \cite{Wilms_2006a} have shown that such a model offers a
good description of pointed \textsl{RXTE} observations of
\mbox{Cyg\,X-1} in all states.

\section{ASM-defined states}\label{sec:states}

\begin{figure}
\begin{minipage}{0.48\textwidth}
\includegraphics[height=\textwidth]{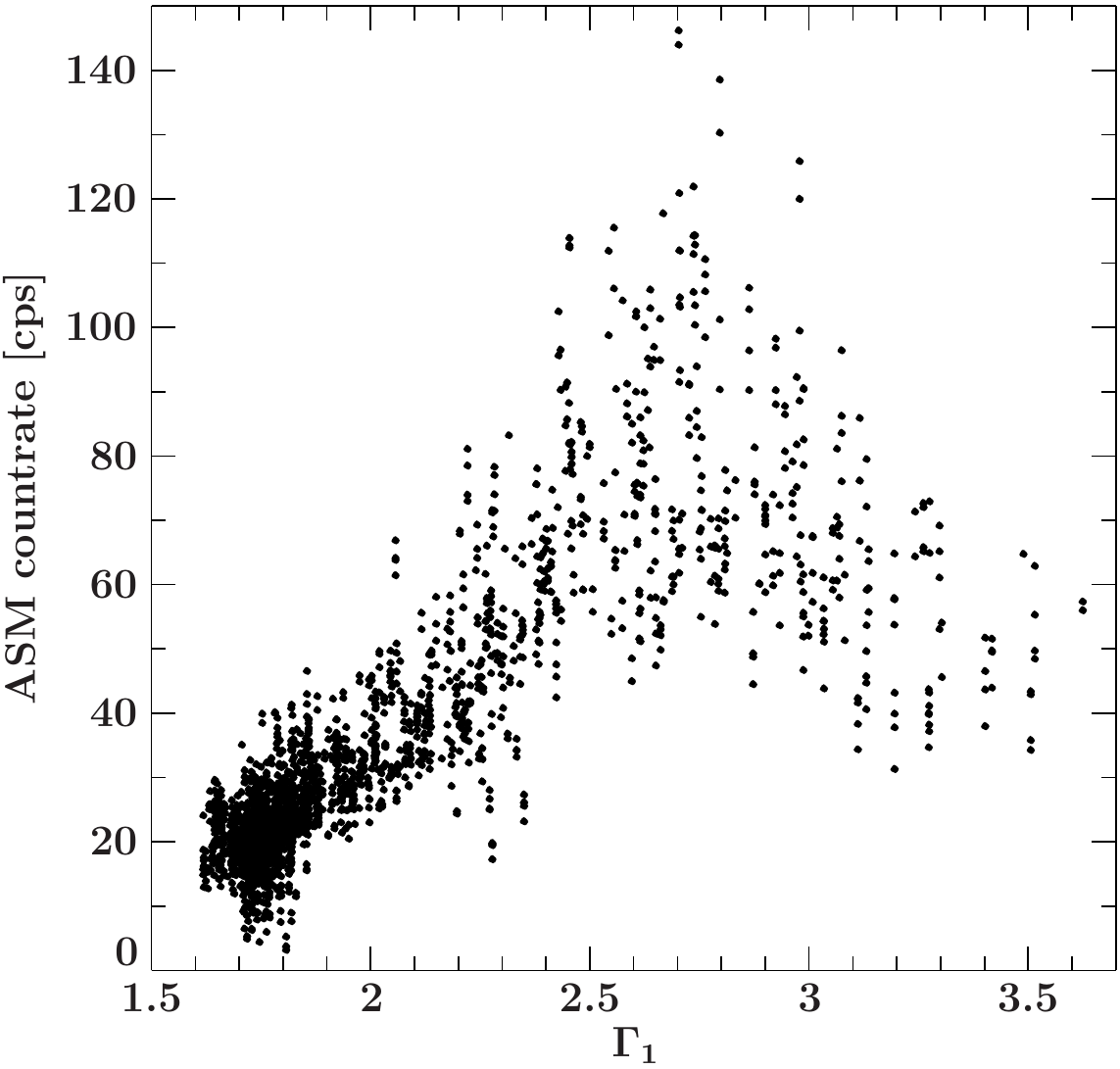}
\end{minipage}\hfill
\begin{minipage}{0.48\textwidth}
\includegraphics[height=\textwidth]{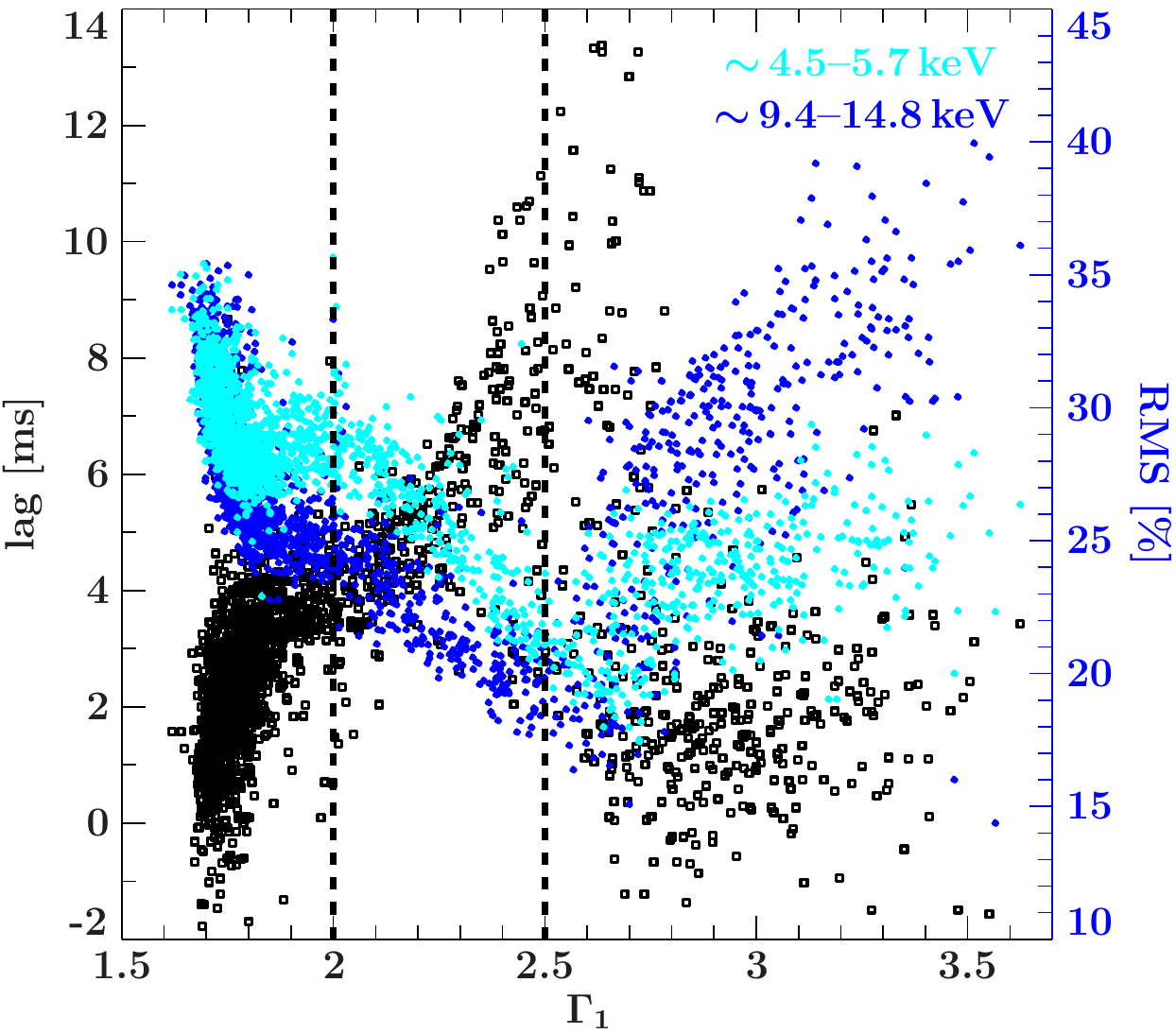}
\end{minipage}
\caption{Dependency of the total ASM count rate (left panel) and X-ray
  timing parameters (time lag between the two bands 4.5--5.7\,keV and
  9.4--14.8\,keV and rms in these two bands, right panel) on the soft
  photon index $\Gamma_1$.}\label{fig:gamma}
\end{figure}

\begin{figure}
\begin{minipage}{0.58\textwidth}
\includegraphics[width=\textwidth]{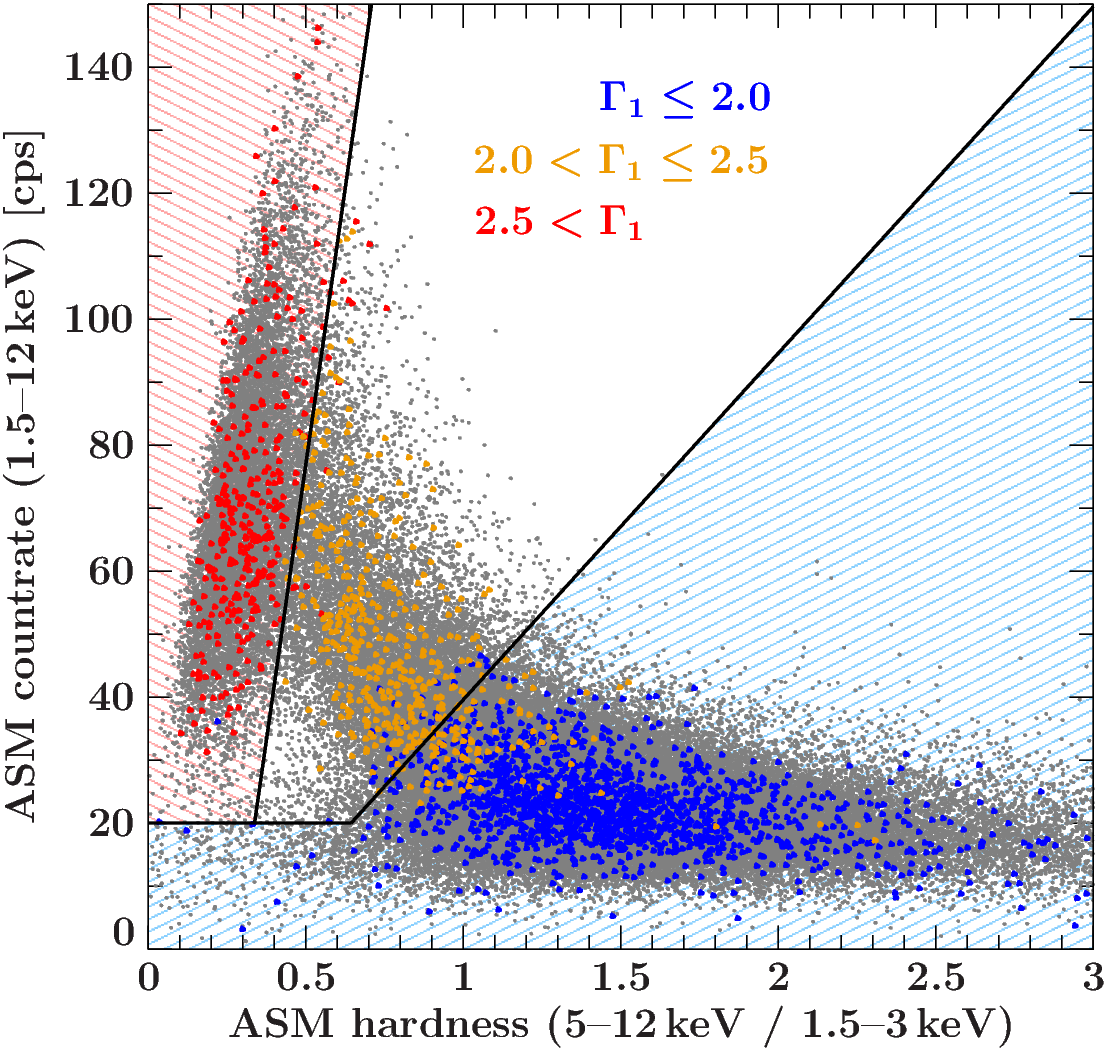}
\end{minipage}\hfill
\begin{minipage}{0.38\textwidth}
  \caption{ASM HID of \mbox{Cyg\,X-1}. Gray data points represent all
    good ASM measurements. ASM measurements which are simultaneous
    with pointed \textsl{RXTE} observations are represented with
    colored points color-coded with the $\Gamma_1$-defined states
    (blue for hard state, orange for intermediate state and red for
    soft state). Cuts between the ASM-defined states are represented
    by black lines. The ASM-defined hard state region is shaded blue,
    the soft state region red, the intermediate state region is shown
    without shading.}\label{fig:map}
\end{minipage}
\end{figure}

As a persistent source \mbox{Cyg\,X-1} does not follow the usual
q-track on the HID, see e.g. \cite{Nowak_2012a} and Fig.~\ref{fig:map}
in this work. The state of the source can best be determined from the
slope of the broad X-ray continuum, by modeling, e.g., an
\textsl{RXTE} spectrum. A strictly simultaneous \textsl{RXTE} spectrum
is, however, not available for an arbitrary observation with a
different instrument, such as \textsl{INTEGRAL}, so that state
determinations for such observations are challenging and so far often
inconclusive.

Usual simple state definitions use the \textsl{RXTE}-ASM data by
setting thresholds in either ASM rate or hardness. For example,
\cite{Wilms_2006a} obtain state estimates by defining the ASM count
rate below 45\,cps as a hard state and above 80\,cps as soft state. We
analyze the 2400 (good, i.e., filtered for negative count rates and
times of instrumental problems) ASM measurements which are strictly
simultaneous to \textsl{RXTE} observations and show that this
approach, while good as a rough guide, fails to account for the
decrease in ASM rate for the softest observations
(Fig.~\ref{fig:gamma}, left panel).

To define \textsl{RXTE} states based on $\Gamma_1$, we consider the
existence of the disk component, which is preferentially required for
modeling those spectra where $ \Gamma_1 > 2.0$, and dependence of the
timing properties of the individual \textsl{RXTE} observations on
$\Gamma_1$. X-ray time lag and fractional rms (Fig.~\ref{fig:gamma},
right panel) are calculated following \cite{Nowak_1999a} using the
same energy and frequency bands as \cite{Boeck_2011a}. We define that
\mbox{Cyg\,X-1} is in hard state if $\Gamma_1 \le 2.0$, in the
intermediate state if $2.0 < \Gamma_1 \le 2.5$ and in the soft state
if $2.5 < \Gamma_1$. The corresponding measurements are shown in color
on the HID in Fig.~\ref{fig:map} over the total ASM measurements shown
in gray. The three $\Gamma_1$-defined states populate different
regions of the total \mbox{Cyg\,X-1} ASM HID.

Finally we define ASM states by introducing cuts in the hardness-rate
space that minimize the contamination of the ASM defined states by
non-corresponding \textsl{RXTE} states. Every measurement with a
count rate $c < 20$\,cps is defined as hard independent of the
hardness $h$. Otherwise the cuts are defined by $c =
m_{\mathrm{hard/soft}} \cdot ( h - h_{0})$, with $h_0 = 0.28$,
$m_{\mathrm{hard}} = 55\,\mathrm{cps}$, $m_{\mathrm{soft }} =
350\,\mathrm{cps}$. The probability of obtaining a misclassification
is below 5\% for the hard, 10\% for the intermediate and 3\% for the
soft state for strictly simultaneous data. A graphical representation
of the cuts is given in Fig.~\ref{fig:map}.

\section{INTEGRAL/IBIS spectra}

\begin{figure}
\begin{minipage}{0.58\textwidth}
\includegraphics[width=\textwidth]{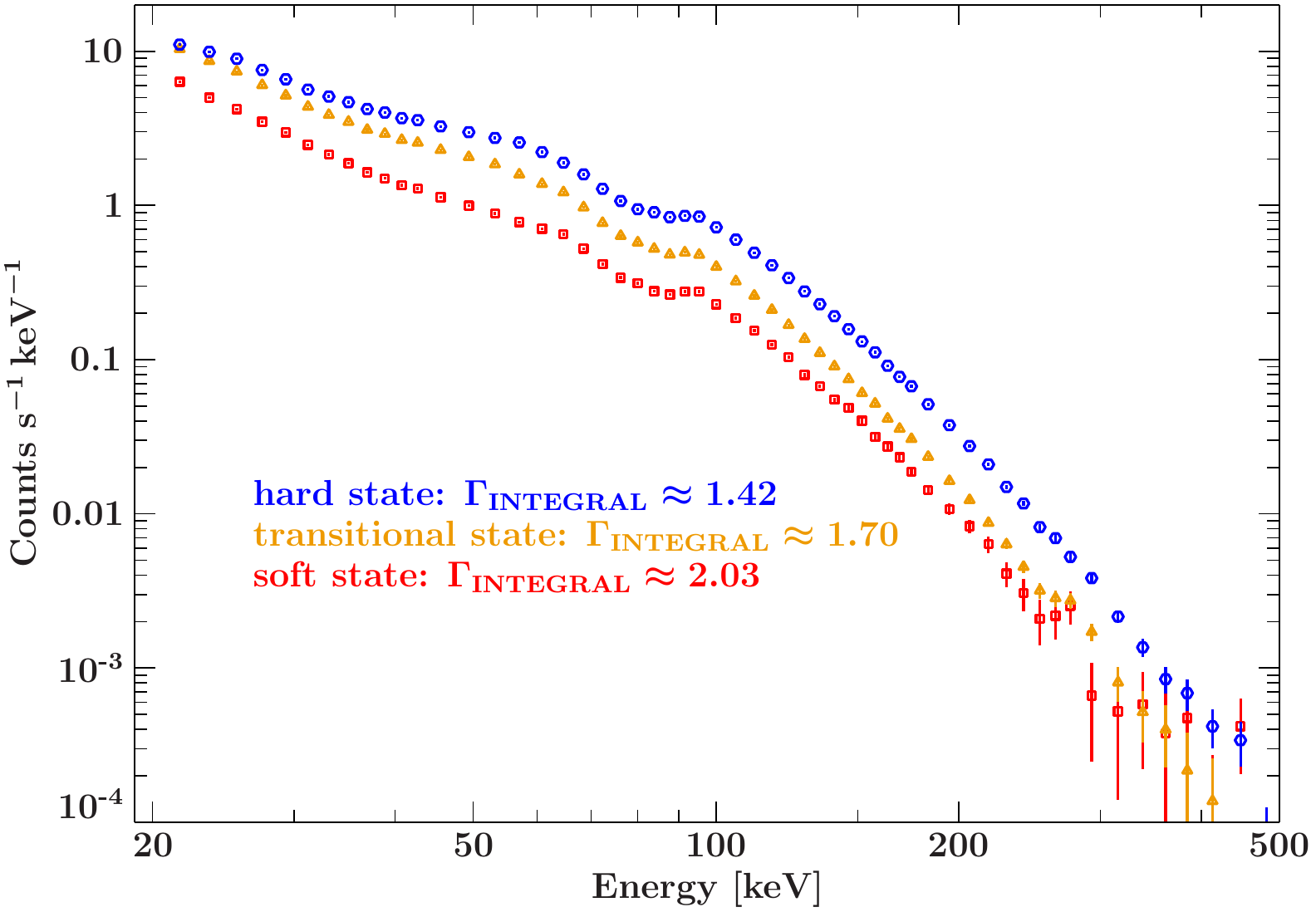}
\end{minipage}\hfill
\begin{minipage}{0.38\textwidth}
  \caption{Preliminary state-resolved \textsl{INTEGRAL}-IBIS spectra
    of \mbox{Cyg\,X-1.}. Note that since \textsl{INTEGRAL}-IBIS data
    only start at $\sim$20\,keV, $\Gamma_{\mathrm{INTEGRAL}}$
    corresponds to the hard photon index $\Gamma_2$ of the broken
    power law models of the \textsl{RXTE} data.}\label{fig:int}
\end{minipage}
\end{figure}

As a proof of concept, we apply a earlier version of the developed ASM
classification on all \textsl{INTEGRAL}-IBIS data of \mbox{Cyg\,X-1}
up to and including 2010. Each individual \textsl{INTEGRAL} science
window is classified by the closest ASM measurement.  We add same
state science windows and obtain state resolved spectra with an
exposure of 1.8\,Ms for the hard state, 1.3\,Ms for the intermediate
state and 0.5\,Ms for the soft state, which we model with a simple
exponentially cut off power law model with photon index
$\Gamma_{\mathrm{INTEGRAL}}$, since \textsl{INTEGRAL}-IBIS data start
well above the spectral break seen in the pointed \textsl{RXTE}
observations at $\sim$10\,keV. Figure~\ref{fig:int} shows that we can
clearly distinguish the different spectral shapes of the three states
with this approach.

\section{Summary and Outlook}

We have developed a novel ASM based classification scheme for the
states of \mbox{Cyg\,X-1} which can be used both for campaigns (e.g.,
with \textsl{INTEGRAL}) and for individual observations (e.g., with
\textsl{XMM}, \textsl{Chandra} or \textsl{Spitzer}). A detailed
discussion of the ASM classification will be presented in an upcoming
paper (Grinberg et al., A\&A submitted), where the feasibility to use
\textsl{Swift}-BAT, MAXI and \textsl{Femi}-GBM data to classify
observations made after the demise of \textsl{RXTE} is also discussed.
We will apply the classification to \textsl{INTEGRAL} data to perform, 
e.g., state-dependent polarization analysis (see \cite{Laurent_2011a}
for the first detection of polarization in \mbox{Cyg\,X-1}).

\acknowledgments 

This work has been partially funded by the Bundesministerium f\"ur
Wirtschaft und Technologie under Deutsches Zentrum f\"ur Luft- und
Raumfahrt Grants 50\,OR\,1007 and 50\,OR\,1113 and by the European
Commission through ITN 215212 ``Black Hole Universe'', was partially
completed by LLNL under Contract DE-AC52-07NA27344. K.P. and
D.M.M. acknowledge support from NASA grant NNX09AT28G for INTEGRAL's
Cycle 7 Guest Observer Programme.  The data analysis presented in this
work was performed with ISIS 1.6.2 \cite{Houck_Denicola_2000a}. We
thank John E. Davis for the development of the \texttt{slxfig} module
used to prepare all figures in this work.


\begin{thebibliography}{99}
\bibitem{Belloni_2010a} Belloni T.M., 2010, \newblock In: {T.~Belloni}
  (ed.) \textsl{Lecture Notes in Physics}, Berlin Springer Verlag,
  Vol. 794. \textsl{Lecture Notes in Physics}, Berlin Springer Verlag,
  p.~53

\bibitem{Boeck_2011a}
{B{\"o}ck} M., {Grinberg} V., {Pottschmidt} K., et~al., 2011, A\&A 533, A8

\bibitem{Fender_2004a}
Fender R.P., Belloni T.M., Gallo E.,  2004, MNRAS 355, 1105

\bibitem{Fender_2009a}
Fender R.P., Homan J., Belloni T.M.,  2009, MNRAS 396, 1370

\bibitem{Houck_Denicola_2000a} {Houck} J.C., {Denicola} L.A., 2000,
  \newblock In: {N.~Manset, C.~Veillet, \& D.~Crabtree} (ed.)
  \textsl{Astronomical Data Analysis Software and Systems IX. ASP
    Conf.}\ Ser.~216, p. 591


\bibitem{Laurent_2011a}
Laurent P., Rodriguez J., Wilms J.,  2011, Science 332, 438

\bibitem{Nowak_1999a}
{Nowak} M.A., {Vaughan} B.A., {Wilms} J., et~al., 1999, ApJ 510, 874

\bibitem{Nowak_2012a}
Nowak M.A., Wilms J., Hanke M., et~al., 2012, Mem. S.A.It. 83, 202

\bibitem{Pottschmidt_2003b}
Pottschmidt K., Wilms J., Nowak M.A., et~al., 2003, A\&A 407, 1039


\bibitem{Wilms_2006a}
{Wilms} J., {Nowak} M.A., {Pottschmidt} K., et~al., 2006, A\&A 447, 245
\end{thebibliography}
\end{document}